%% file: 00main.tex
\documentclass[sigconf]{acmart}
\AtBeginDocument{%
  \providecommand\BibTeX{{%
    \normalfont B\kern-0.5em{\scshape i\kern-0.25em b}\kern-0.8em\TeX}}}

\copyrightyear{2022}
\acmYear{2022}
\setcopyright{acmcopyright}
\acmConference[ICSE-NIER'22]{New Ideas and Emerging Results }{May 21--29, 2022}{Pittsburgh, PA, USA}
\acmBooktitle{New Ideas and Emerging Results (ICSE-NIER'22), May 21--29, 2022, Pittsburgh, PA, USA}
\acmPrice{15.00}
\acmDOI{10.1145/3510455.3512790}
\acmISBN{978-1-4503-9224-2/22/05}

\usepackage{float}
\floatstyle{plaintop}
\restylefloat{table}

\usepackage{url}
\usepackage{multirow}
\usepackage[most]{tcolorbox}

\begin{document}

\title{Evaluating Commit Message Generation: To BLEU Or Not To BLEU?}

\author{Samanta Dey}
\email{samanta@cmi.ac.in}
\author{Venkatesh Vinayakarao}
\email{venkateshv@cmi.ac.in}
\affiliation{%
  \institution{Chennai Mathematical Institute}
  \city{Chennai}
  \country{India}
}

\author{Monika Gupta}
\email{mongup20@in.ibm.com}
\author{Sampath Dechu}
\email{sampath.dechu@in.ibm.com}
\affiliation{%
  \institution{IBM Research}
  \city{Delhi}
  \country{India}
}

\renewcommand{\shortauthors}{Samanta et al.}

\input{01abstract}



\maketitle

\input{01introduction}

\input{02background}

\input{03experimentalsetup}

\input{04results}

\input{05thelogmnextmetric}
\input{06relatedwork}

\input{07futurework}

\input{08conclusion}

\clearpage

\bibliographystyle{ACM-Reference-Format}
\bibliography{09references}

\end{document}

%% file: 01abstract.tex
%Abstract
%Intro
%What is LMG? WHy should we look at the evaluation for LMG?
%Factors
%We need to use the right metric. 
%So, we ask the following research questions
%What factors...
%RQ2
%RQ3

%Background

%Experimental Setup and Results

%Related Work
%Conclusion and Future Work

\begin{abstract}

Commit messages play an important role in several software engineering tasks such as program comprehension and understanding program evolution. However, programmers neglect to write good commit messages. Hence, several Commit Message Generation (CMG) tools have been proposed. We observe that the recent state of the art CMG tools use simple and easy to compute automated evaluation metrics such as BLEU4 or its variants. The advances in the field of Machine Translation (MT) indicate several weaknesses of BLEU4 and its variants. They also propose several other metrics for evaluating Natural Language Generation (NLG) tools. In this work, we discuss the suitability of various MT metrics for the CMG task. Based on the insights from our experiments, we propose a new variant specifically for evaluating the CMG task. We re-evaluate the state of the art CMG tools on our new metric. We believe that our work fixes an important gap that exists in the understanding of evaluation metrics for CMG research. 

\end{abstract}

%% file: 01introduction.tex
\section{Introduction}
\label{sec:introduction}
Commit messages are Natural Language (NL) descriptions of code changes. Developers submit such textual description along with their code changes. These messages, called as \textit{commit messages} or \textit{log messages}, are important to the overall software development process and reduce the time taken to understand the code change. Often, developers provide low quality commit messages or even no message. Dyer et al. [16] analyzed  23,000 Java projects on SourceForge\footnote{https://sourceforge.net/} and reported that around 14\% of commit messages were empty. Therefore, CMG tools are required. Commit Message Generation (CMG) tools take code changes as input and produce commit messages as output. 

Automatically generated commit messages are evaluated against the ground-truth. Human evaluation would provide the best assessments, but it is expensive and time-consuming. Automated evaluation of the CMG task so far has been limited to reusing metrics popularly used for various Machine Translation (MT) and Natural Language Generation (NLG) tasks. 

%A metric for automatically evaluating CMG tools should have high correlation with such human provided commit messages. It should also be able to capture the semantics of the code changes.

The first known automatic evaluation metric BLEU~\cite{10.3115/1073083.1073135} relied on n-gram precision. BLEU computes word overlap between the predicted  and the reference gold-standard written by the project members. METEOR~\cite{banerjee2005meteor} is another metric which is based on unigram matches on the words and also their stems with additional synonyms database. In addition to these, numerous other metrics~\cite{sai2020survey} have evolved for automatically evaluating Machine Translation (MT) systems. In addition to semantic scoring through word overlap, these metrics propose different ways to penalize the length, and score word order alignment. 

BLEU and its variants may not be suitable for assessing commit messages. Reiter~\cite{reiter2018a} claims that the assumption of word overlap correlating with real-world utility needs to be validated through user studies or task performance. %For example, BLEU-4, the variant used in ~\cite{hoang2020cc2vec} gives a score of zero for the commit message pair, `changed path to libraries' and `updated libraries' even though they mean the same. 
It is unclear, which metric is most appropriate for evaluating the CMG task. 

% Although these many metrics exist, BLEU-4 has been most widely used by several recently proposed CMG tools including CC2Vec~\cite{10.1145/3377811.3380361}, NMT~\cite{10.1145/3238147.3238190} and NNGen~\cite{NNGen}.

%Typical metrics used are BLEU-4~\cite{} and its variants, METEOR~\cite{} and NIST~\cite{}. We observe that there are serious limitations concering their applicability to Commit Message Generation (CMG) task of Software Engineering (SE). We propose a new metric for automatic evaluation of CMG tools.

%which employs several augmented factors, after careful exploration of some common and uncommon Evaluation Metrics.
%METEOR proposed an alternative way of calculating matched chunks to describe the n-gram matching degree between predictions and references. 

%We observe serious limitations in these metrics due to the way they handle semantic scoring, length penalties and word alignment. Hence, we propose a new metric specifically designed for CMG task.

The fundamental factors behind the widely used MT based metrics are \textit{semantic scoring}, \textit{length penalty} and \textit{word order alignment}. In addition, Tao et al.~\cite{tao2021evaluation} observe that \textit{case sensitivity} and \textit{smoothing} could be potential factors affecting CMG evaluation. %Section~\ref{sec:ourapproach} explains these factors in detail. 
The relevance of these factors with respect to commit messages have not been studied. Understanding their effect on commit message quality as per expert perception i.e., as per human evaluation will help us in deciding the metric to use for evaluating the commit messages. Hence, we ask the following research questions.

\begin{enumerate}
    \item[RQ1] What factors affect commit message quality as per expert perception?
    \item[RQ2] Which metric is best suited to evaluate commit messages?
    \item[RQ3] How do the CMG tools perform on the new metric?
\end{enumerate}

To the best of our knowledge, this is the first work to evaluate the influence of various factors on existing metrics, and also propose a metric specifically to evaluate CMG tools.

%Reiter~\cite{reiter2018a} claims that the assumption of word overlap correlating with real-world utility needs to be validated through user studies or task performance. 

%% file: 02background.tex
\section{Background}
\label{sec:background}

\subsection{Commit Message Generation Tools}
\label{sec:BGonCMG}
Several CMG tools have been proposed.Jiang et al.~\cite{jiang2017automatically} proposed CommitGen which uses an attentional Recurrent Neural Network (RNN) encoder-decoder based Neural Machine Translation (NMT) model \textit{Nematus} to translate code change diffs into commit messages. The CommitGen model is trained using a corpus of code changes and human-written commit messages from the top 1k Github projects. NMT proposed by Loyola~\cite{loyola2017neural} in 2017, is similar to CommitGen, but it is guided by a global attention model proposed by Luong et al.~\cite{luong2015effective} instead of the Bahdanau attention model in NMT. This model is trained using code changes and human-written commit messages from  popular Github projects across multiple programming languages. NNGen proposed by Liu et al.~\cite{NNGen} in 2018, is a retrieval-based model which is based on the Nearest Neighbour and Bag-of-Words (BOW) approach. The NNGen model is trained on a filtered version of the CommitGen dataset.

\subsection{Automated Evaluation in Machine Translation}
\label{sec:BGonAEinMT}
Several automated metrics have been proposed~\cite{chauhan2021adableu}. The prominent MT evaluation metrics used in evaluating CMG tools can be broadly categorised as a) Precision based, b) Recall based, c) F-Score based, and d) Edit Distance based metrics.

\paragraph{Precision based Metrics} During word matches, precision based metrics score higher if there are more matches for words in the predicted text. BLEU4~\cite{papineni2002bleu} is a classic example of a precision based metric. BLEU4 is calculated as shown in Equation~\ref{eq:bleu4} where BP is the brevity penalty, $w_k$ refers to the empirically chosen weight of each k-gram and $p_k$ is the match score for the k-gram. 
\begin{equation}
    \text{BLEU4}=\text{BP} \cdot exp \left( \displaystyle  \sum_{k=1}^{4} w_k \cdot log(p_k)\right)
    \label{eq:bleu4}
\end{equation}
There are several variants of BLEU such as BLEUMoses~\cite{koehn2007moses}, BLEUNorm~\cite{loyola2017neural} and BLEUCC~\cite{chen2014systematic}. BLEUNorm applies smoothing to the match score as $p_k = \frac{m_k+1}{l_k+1}$ where $m_k$ is  number of matched k-grams between the reference and the predicted text and $l_k$ is the the total number of
k-grams in the predicted text. BLEUCC makes further improvisation on this by applying the assumption that a k-gram match score can be smoothed by k-1 and k+1 match scores i.e., $m_k = \frac{m_{k-1} + m_k + m_{k+1}}{3}$. 

\paragraph{Recall based Metrics} During word matches, recall based metrics score higher if more words are retrieved considering each word in the reference text. ROUGE~\cite{lin2004rouge} is an example of a recall based metric, which is calculated as shown in Equation~\ref{eq:rouge_n},  where $\text{Count}_{\text{match}}$ is the number of n-gram matches between predicted and reference (Ref) text while $\text{Count}({\text{n-gram})}$ refers to the number of n-grams.
\begin{equation}
        \text{ROUGE-n}=\frac{\sum_{\text{n-grams} \in \text{Ref}} \text{Count}_{\text{match}}(\text{n-gram})}{\sum_{\text{n-grams} \in \text{Ref}} \text{Count}(\text{n-gram})}
        \label{eq:rouge_n}
\end{equation}
     
%\paragraph{F-Score based Metrics} F-Score is the harmonic mean of both precision and recall. F-Score based metrics balance the trade-off between the need for both precision and recall. METEOR~\cite{banerjee2005meteor} is an example for F-Score based metric which, in addition to exact word matches, also performs stem and synonym matches. The F-score of METEOR is calculated as shown in Equation~\ref{fscore1}.
%\begin{equation}
            %\text{F-Score}=\frac{\text{10PR}}{\text{R + 9P}}
            %\label{fscore1}
%\end{equation}
%where, P (Precision) is the ratio of the number of matched unigrams to the length of predicted text and R (Recall) is the ratio of the number of matched unigrams to the length of reference text. 
\paragraph{F-Score based Metrics} METEOR is an F-Score based metric. It considers a fragmentation penalty, calculated as $\text{Frag Penalty}=0.5 * \left[ \frac{\text{\# chunks}}{\text{\# unigrams matched}} \right]^3$ where chunks are contiguous predicted unigrams mapped to contiguous unigrams in the reference. The final METEOR score is thus calculated using Equation~\ref{meteor}.
\begin{equation}
             \text{METEOR} = \text{F-score} * {(1-\text{Frag Penalty})}
            \label{meteor}
\end{equation}

METEOR-NEXT~\cite{denkowski:lavie:meteor-wmt:2010} is a hyper parametrized version of METEOR.

\paragraph{Edit Distance based Metrics} In Edit Distance based metrics, number of word operations such as insertions, deletions and substitutions are used to compute the word edit distance between the given and reference texts. TER~\cite{snover2006study} is an example. The TER metric score is calculated as shown in Equation~\ref{ter}.
 \begin{equation}
        \text{TER}=\frac{\text{\#substitutions + \#deletions + \#insertions}}{\text{\#words in reference text}}
        \label{ter}
    \end{equation}

%\paragraph{Other Metrics} BERT regressor~\cite{shimanaka2019machine} is a metric based on BERT encoder. Takahashi et al.~\cite{takahashi2020automatic} propose a metric trained on cross-lingual language models. The BERT model \textit{RoBERTa}~\cite{liu2019roberta}, based on Transformers, can be used to find the semantic similarity between two text pieces.The semantic similarity score for examples in Table~\ref{tab:metrics} are given.

%\subsection{Automated Evaluation in LMG Tools}
%\label{sec:BGonAEinLMGTools}

%% file: 03experimentalsetup.tex
\section{Experimental Setup}
\label{sec:experimentalsetup}

\input{tblCorrelationFactors}

%What resources do we exploit to conduct the experiments? (Human evaluations dataset, metrics considered, CMG tools, dataset used)
%RQ1 & RQ2
\subsection{Selection of Factors}

Commit messages although defined as NL descriptions of code changes, are different from normal NL text. To address RQ1, we gather several factors from the various evaluation metrics and related literature.

Papineni et al.~\cite{papineni2002bleu} propose that a good evaluation metric should ensure that the predicted text be neither too long nor too short. Hence precision based metrics such as BLEU~\cite{papineni2002bleu} and its variants use length penalizers to penalize shorter predictions. “The gunman killed the cop” and “The cop killed the gunman” are not the same sentences. In order to handle ordering, a simple bag of words representation is not suitable. Metrics such as METEOR~\cite{banerjee2005meteor} and METEOR-NEXT~\cite{denkowski:lavie:meteor-wmt:2010} account for the alignment of matched words in the predicted and reference sentences. The pair “Update change" and “Updated changes" essentially convey the same meaning, although exact-word matchers in BLEU and ROUGE would give a score of zero instead of one. Hence, in addition to exact matches, stemmed and paraphrased matchings are also considered by METEOR and METEOR-NEXT. Banerjee and Lavie~\cite{banerjee2005meteor} convert text to lower case as a part of preprocessing in the implementation of METEOR. Punctuations are often treated as separate words by the default parsers and matchers in various programming languages which heavily affects the scores. %Consider the difference in use of punctuation in the pair "py : use dict getter to retrieve binary" and "[ py ] fix binary capabilities" which affects evaluation. 
BLEU4 geometrically averages the n-gram matches, thereby, giving zero score for pairs like "added chain of responsibility class diagram" and "added chain diagram" due to missing n-gram matches of order three or above. BLEUCC and BLEUNorm are smoothed versions of BLEU. Therefore, we consider the factors namely length, word order, semantics, case, punctuation and smoothing as shown in Table~\ref{tab:factors_working}.

\subsection{Comparison with Human Annotations}

We use a dataset of 100 human evaluated reference and predicted commit message pairs shared by Tao et al.~\cite{tao2021evaluation}. The data is manually labelled between 0 to 4 by three domain experts and validated for reliability. We  took the arithmetic mean to obtain a single averaged score of human evaluation for each pair of reference and predicted commit messages. In our experiments, we compare the scores produced using the MT metrics discussed in Section~\ref{sec:BGonAEinMT} with these human annotation scores using Spearman's correlation~\cite{zar2005spearman}. We do it once in the presence of the factor and once in its absence. Table~\ref{tab:factors_working} lists the modifications in the metric formulae due to inclusion and exclusion of the various factors used in our study. %In Table~\ref{tab:factors_working}, $\text{F-score}_{no}$ and $\text{F-score}_{ns}$ respectively denote the modified F-scores without considering word order and semantics.

%To observe the effect of the factor \textit{length}, we consider the metrics BLEU and its variants. We obtain the correlation with human evaluation scores, firstly by removing the brevity penalty factor `BP' and then by replacing it back. For other metrics such as METEOR and ROUGE which do not apply length penalty, the formulation remains the same. Likewise, since METEOR and METEOR-NEXT incorporate the \textit{word order} or \textit{alignment} factor, we first calculate the modified `$\text{F-score}_{no}$' without considering alignment in the formulation and then the original scores. A similar approach is followed in case of \textit{semantic scoring} by calculating a modified `$\text{F-score}_{ns}$' factor for the formula without semantics considered. For the factors \textit{case-folding} and \textit{punctuation removal}, we re-calculate all the metrics by adding preprocessing steps for lower-casing and removing punctuation using the modified formulae, denoted by `$\text{Metric Formula}_{cf}$' and `$\text{Metric Formula}_{pr}'$ respectively in Table~\ref{tab:factors_working}. Finally, for the smoothing factor, we first consider the simple BLEU4 metric which applies no smoothing and then its smoothed variants BLEUNorm and BLEUCC by modifying `$p_k$' to `$p_{ks}$' in the formulation of BLEU. 
%RQ3
\subsection{CMG Tools, Metrics and the Commit Dataset}
To answer RQ3, we consider the CMG tools discussed in Section~\ref{sec:BGonCMG}. Different CMG tools use different datasets for their evaluation. In order to compare them, we need a unified dataset. NMT and NNGen use datasets consisting of only Java code changes that are relatively small (i.e., not exceeding 50K). We follow Tao et al.~\cite{tao2021evaluation} and use their Multi-Language Commit Message (MCMD) dataset. This dataset has 3.6M commit messages in five popular Programming Languages (PLs) from the top 100 starred projects on GitHub.

\input{tblFactors}

%% file: tblCorrelationFactors.tex
\begin{table*}[t]
\small
\centering
\begin{tabular}{|c|c|c|c|c|c|c|c|c|c|c|c|c|c|}
\hline
\multirow{2}{*}{Metric} & \multicolumn{2}{c|}{Length} & \multicolumn{2}{c|}{Word Order} & \multicolumn{2}{c|}{Semantics} & \multicolumn{2}{c|}{Case Folding} & \multicolumn{2}{c|}{Punctuation} & \multicolumn{2}{c|}{Smoothing} & \multirow{2}{*}{Clean}\\ \cline{2-13} 
                        & Without        & With       & Without          & With        & Without         & With         & Without           & With          & Without              & With              & Without         & With &        \\ \hline
BLEU4                                       & 0.69                        & 0.705                    & 0.705                       & 0.705                    & 0.705                       & 0.705                    & 0.705                       & 0.717                    & 0.705                       & 0.707                    & 0.705                       &  0.691, 0.681 & 0.705\\
BLEUNorm                                    & 0.683                       & 0.691                    & 0.691                       & 0.691                    & 0.691                       & 0.691                    & 0.691                       & 0.703                    & 0.691                       & 0.699                    & 0.691                       & 0.691                      & 0.691     \\
BLEUCC                                      & 0.683                       & 0.681                    & 0.681                       & 0.681                    & 0.681                       & 0.681                    & 0.681                       & 0.691                    & 0.681                       & 0.693                    & 0.681                       & 0.681                       &0.681    \\
METEOR                                      & 0.748                       & 0.748                    & 0.725                       & 0.748                    & 0.707                       & 0.748                    & 0.74                        & 0.748                    & 0.748                       & 0.807                    & 0.748                       & 0.748                      &0.748     \\
METEOR-NEXT                                 & 0.761                       & 0.761                    & 0.756                       & 0.761                    & 0.722                       & 0.761                    & 0.736                       & 0.761                    & 0.761                       & 0.822                    & 0.761                       & 0.761                      & 0.761     \\
ROUGE1                                      & 0.723                       & 0.723                    & 0.723                       & 0.723                    & 0.723                       & 0.723                    & 0.723                       & 0.796                    & 0.723                       & 0.781                    & 0.723                       & 0.723                      & 0.723     \\
ROUGE2                                      & 0.443                       & 0.443                    & 0.443                       & 0.443                    & 0.443                       & 0.443                    & 0.443                       & 0.485                    & 0.443                       & 0.485                    & 0.443                       & 0.443                       & 0.443    \\
ROUGEL                                      & 0.728                       & 0.728                    & 0.728                       & 0.728                    & 0.728                       & 0.728                    & 0.728                       & 0.799                    & 0.728                       & 0.781                    & 0.728                       & 0.728                       & 0.728    \\
TER                                         & 0.568                       & 0.568                    & 0.568                       & 0.568                    & 0.568                       & 0.568                    & 0.568                       & 0.583                    & 0.568                       & 0.54                     & 0.568                       & 0.568                      & 0.568 \\      \hline                
\end{tabular}
\caption{Spearman's correlations of metrics with human judgments w.r.t. potential factors. ``Clean'' refers to the correlation of metric in its original formulation without any changes.}
\label{tab:correlation_factors}
\end{table*}

%% file: tblFactors.tex
% \usepackage{multirow}
\begin{table}[t]
\small
\begin{tabular}{|p{1.3cm}|p{0.8cm}|p{2.4cm}|p{3cm}|}
\hline
\multirow{2}{*}{{Factor}} & \multicolumn{1}{c|}{\multirow{2}{*}{{Metrics}}} & \multicolumn{2}{c|}{{Formula}} \\ \cline{3-4} 
                                 & \multicolumn{1}{c|}{}                                 & {Factor not included}    & {Factor included}   \\ \hline
\multirow{2}{*}{Length}          & B4, BN, BCC                               &           $  exp \left( \displaystyle  \sum_{k=1}^{4} w_k log(p_k)\right)$          &        $\text{BP} * exp \left( \displaystyle  \sum_{k=1}^{4} w_k log(p_k)\right) $        \\ \cline{2-4} 
                                 & Others                                         & \multicolumn{2}{c|}{No Change.}        \\ \hline
\multirow{2}{*}{Word Order}      & M, MN                                &    ${\text{F-score}_{\text{no alignment}}} * {(1-\text{FP})} $                   &   $\text{F-score} * {(1-\text{FP})} $                \\ \cline{2-4} 
                                 & Others                                         & \multicolumn{2}{c|}{No Change.}        \\ \hline
\multirow{2}{*}{Semantics}       & M, MN                                   &  ${\text{F-score}_{\text{exact matches}}} * {(1-\text{FP})} $                    &     $\text{F-score} * {(1-\text{FP})} $             \\ \cline{2-4} 
                                 & Others                                         & \multicolumn{2}{c|}{No Change.}        \\ \hline
Case                    & All                                           & No case folding.     &  Use lower case text.             \\ \hline
Punctuation              & All                                           & No Change.      &  Remove punctuations.                \\ \hline
\multirow{2}{*}{Smoothing}       & B4                                                 &       No Change.            &    BN, BCC             \\ \cline{2-4} 
                                 & Others                                         & \multicolumn{2}{c|}{No Change.}        \\ \hline
\end{tabular}
\caption{Computation after addition/removal of factors in metrics. Here, B4 is BLEU4, BN is BLEUNorm, BCC is BLEUCC, M is METEOR and MN is METEORNEXT, FP is Fragmentation Penalty. }
\label{tab:factors_working}
\end{table}

%$ exp \left( \displaystyle  \sum_{k=1}^{4} w_k log(p_k)\right)$  

%$ exp \left( \displaystyle  \sum_{k=1}^{4} w_k log(\frac{m_k}{l_k})\right)$

%% file: 04results.tex
\section{Results}

\subsection{RQ1: What factors affect commit message generation as per expert perception?}

Table~\ref{tab:correlation_factors} shows the Spearman's correlation values with and without the six factors applied to the various metrics. A factor is assumed to affect the evaluation of CMG if its presence in a metric increases the Spearman's correlation with human evaluation. The observations from the Table~\ref{tab:correlation_factors} indicate that the factors \textit{Length, Word Alignment, Semantic Scoring, Lower-Casing} and \textit{Punctuation Removal}, when incorporated in the corresponding metrics, improve correlation with human judgements. While inclusion of the factor \textit{Smoothing} reduces the correlation with human evaluation.

%A point to be noted here is that, since both BLEUCC and BLEUNorm are smoothed versions of BLEU4, we have two correlation values (for BLEUNorm and BLEUCC respectively) in the last column of the 1st row in Table~\ref{tab:correlation_factors}. 
\vspace{5pt}
\begin{tcolorbox}[colback=gray!5!white,colframe=gray!75!black]
Commit message quality is influenced by \textbf{Length, Word-Alignment, Semantic Scoring, Lower-Casing} and \textbf{Punctuation-Removal}. These factors should be considered in a metric designed for the purpose of CMG evaluation.
\end{tcolorbox}.

\subsection{RQ2: Which metric is best suited to evaluate commit messages?}
\input{tblPerformModels}
None of the standard MT metrics discussed in Section~\ref{sec:BGonAEinMT} have all the affecting factors incorporated in them. This motivates the construction of a new metric. We propose a modified version of the METEOR-NEXT metric, called \textit{Log-MNEXT}. Section~\ref{sec:OurApproach} discusses the construction of \textit{Log-MNEXT}. To validate the goodness of \textit{Log-MNEXT} to the existing MT metrics, its Spearman's correlation with human evaluation score is calculated and compared with that of the other metrics in Table~\ref{tab:correlation_factors}. 
\vspace{5pt}
\begin{tcolorbox}[colback=gray!5!white,colframe=gray!75!black]
\textit{Log-MNEXT} has the highest correlation with human evaluation as compared to the standard MT evaluation metrics with a correlation value of 0.831.
\end{tcolorbox}.
\subsection{RQ3: How do the CMG tools perform on the new metric?}

The CMG models discussed in Section~\ref{sec:BGonCMG} are evaluated using variants of BLEU. In a similar experimental setup as that of Tao et al.~\cite{tao2021evaluation}, we compare the performances of CommitGen, NMT and NNGen using \textit{Log-MNEXT} metric.

The MCMD dataset is split across various programming languages (PLs). The average \textit{Log-MNEXT} scores in percentages, across datasets of different PLs for each of the models, is given in Table~\ref{tab:perform_models}. Overall, the retrieval-based model NNGen outperforms CommitGen and NMT, with an average \textit{Log-MNEXT} score of 14.13. % As per Tao et al.~\cite{sai2020survey}, a probable reason of NNGen getting a high score in the majority of dataset varieties is the duplication of commit messages.

\vspace{5pt}
\begin{tcolorbox}[colback=gray!5!white,colframe=gray!75!black]
NNGen performs the best when evaluated on the MCMD dataset using the \textit{Log-MNEXT} metric.
\end{tcolorbox}

%% file: tblPerformModels.tex
\begin{table}[t]
\small
\vspace{-5pt}
\begin{tabular}{|c|l|l|l|l|l|l|}
\hline
\multirow{2}{*}{Model} & \multicolumn{6}{c|}{$\text{MCMD}_{\text{data}}$}                             \\ \cline{2-7} 
                       & C++   & C\#   & Java & JS    & Py    & Avg            \\ \hline
CommitGen              & 11.94 & \textbf{18.35} & 9.63 & 18.11 & 8.27  & 13.26          \\ \hline
NMT                    & 11.69 & 17.96 & \textbf{10.7} & 13.58 & 9.34  & 12.65          \\ \hline
NNGen                  & \textbf{13.82} & 16.89 & 7.4  & \textbf{18.25} & \textbf{14.27} & \textbf{14.13} \\ \hline
\end{tabular}
\caption{Performances of CMG models.}
\label{tab:perform_models}
\end{table}

%% file: 05thelogmnextmetric.tex
\section{Log-MNEXT Metric}
\label{sec:OurApproach}

%\paragraph{Length} The need for length penalty arises when we want to penalize predictions either shorter (as in precision-based metrics like BLEU) or longer than the reference. Recall and precision respectively account for length variations due to excess words in the prediction and the reference. Since METEOR-NEXT combines both precision and recall into a single F-score, it already penalizes length variations. Thus, a separate length penalty factor seems redundant in this case.

%\paragraph{Word Alignment} \textit{Log-MNEXT} defines an alignment as a mapping between matched unigrams, such that a single unigram in one string cannot map to more than one unigram in the other string. For two alignments with the same number of mappings, the one is chosen with the fewest crosses, that is, with fewer intersections of two mappings. Banerjee and Lavie~\cite{banerjee2005meteor} discuss in detail how the alignment works.

METEOR-NEXT incorporates most of the relevant factors. Hence, we base \textit{Log-MNEXT} on METEOR-NEXT. In addition to exact word matchings, \textit{Log-MNEXT} also performs stemmed and paraphrased matchings as in METEOR and METEOR-NEXT metrics. \textit{Log-MNEXT} performs string lower-casing and removes punctuation as a preprocessing step, before looking for word matches or performing word alignment between the predicted and reference texts. 

Consider the case where we have identical reference and predicted messages. Any human annotator will not penalize this. However, the denominator of the fragmentation penalty factor \textit{Frag Penalty} of METEOR-NEXT given by Equation~\ref{fragpen} depends on the number of unigram matches. Hence, it penalizes this case. \textit{Log-MNEXT} improvises on the \textit{Frag Penalty} by assigning no penalty score in such cases.
\begin{equation}
            \textit{Frag Penalty}=
            \begin{cases}
            0   & \text{if} ~\textit{Ref} \equiv \textit{Pred}\\
			\beta * \left[ \frac{\text{\# chunks}}{\text{\# unigrams matched}} \right]^\gamma &\text{otherwise}
		    \end{cases}
            \label{fragpen}
            \end{equation}
The \textit{Log-MNEXT} score is finally obtained using equation~\ref{logMN}.
\begin{equation}
            {\textit{Log-MNEXT} = \textit{F-score * (1-\textit{Frag Penalty})}}
            \label{logMN}
\end{equation}

To get the F-Score, Precision (P) and recall (R) are calculated by assigning weights to the various unigram matches.
\begin{equation}
P = \frac{\sum w_im_i}{L_{\text{pred}}}
\text{~and~}
R = \frac{\sum w_im_i}{L_{\text{ref}}}
\end{equation}
where for a matcher type i $\in$ \{exact match, stemmed match, synonym match\}, $w_i$ is the weight and $m_i$ is the number of matched unigrams. $L_{\text{pred}}$ is the length of predicted text and $L_{\text{ref}}$ is the length of reference text. The weighted \textit{F-score} with $0 < \alpha < 1$ is calculated as $\frac{PR}{\alpha P + (1-\alpha) R}$. For the optimum values of $\alpha$,  $\beta$ and $\gamma$, we use the values suggested by Denkowski and Lavie~\cite{denkowski:lavie:meteor-wmt:2010}. 

%% file: 06relatedwork.tex
\section{Related Work}
\label{sec:relatedwork}
There are works comparing~\cite{han2016machine, mathur-etal-2020-tangled,sai2020survey,celikyilmaz2020evaluation} automated evaluation metrics for machine translation. Our work is different from them because we discuss the suitability of metrics used in evaluating commit message generation tools. Tao et al.'s~\cite{tao2021evaluation} work is the closest to our work. However, they limit their study to BLEU variants. They report that the models rank inconsistently across the different metrics. They find that BLEUNorm is most correlated to human annotations. They suggest smoothing and case sensitivity as two potential reasons for it to perform well. Finally, they report existence of heavy sensitivity of these metrics on the dataset. These are related and relevant to our work. However, we consider other popular metrics from MT and NLG for evaluation. We find that a new metric is necessary. Further, we propose the new metric and evaluate the performance of the existing models on the new metric.
%For example, BLEU-Norm drops from 34.74 to 9.07 when dataset is changed from CommitGen to CoDiSum.   
%Firstly, they report large variance among the BLEU scores. They compared the semantic scores provided by three human evaluators in a scale of 0 to 4, to the scores of various BLEU variants. Secondly, t

%% file: 07futurework.tex
\section{Future Work}
The results of our experiments emphasize on the use of the novel metric \textit{Log-MNEXT} for evaluating commit messages. We now enlist key future directions.

\paragraph{Building larger human annotated dataset} For the purpose of this work, we have used existing human annotated dataset. The size of the dataset (100) leads to internal validity. We plan to build our own larger dataset, containing more number of reference and predicted commit message pairs. Claims made on the basis of such a dataset are expected to be more sound and strong.

\paragraph{Working on learning based metrics} This work limits to non-learning based i.e., simple and fast metrics such as BLEU. This may lead to external validity. This is mitigated to a large extent since the existing CMG tools use the fast metrics for evaluation for various reasons including the fact that even the learning based metrics have not been established to correlate significantly with human annotations better than the fast metrics. We plan to study learning based metrics in detail as future work.

\paragraph{Revising the Log-MNEXT metric} This work presents the most basic form of the metric Log-MNEXT. A major portion in the construction of the metric has been adapted from the existing METEOR-NEXT metric. Improvisations in the form of better synonym and paraphrase matching, cautious removal of punctuation from sentence pairs and careful case-folding before evaluation, are some important scopes of future work.

\paragraph{Building an exhaustive commit dataset} To compare the performances of the CMG tools on the Log-MNEXT metric, the MCMD dataset has been used. Although the language diversity in MCMD has been expanded to 5 popular PLs, it is still not exhaustive. Building an exhaustive dataset by including more PLs, is a direction of future work. However, appropriate caution is needed when generalizing our findings across PLs.

%We have used existing human annotated dataset. The size of the dataset (100) leads to internal validity. We plan to build a larger dataset. This work limits to non-learning based i.e., simple and fast metrics such as BLEU. This may lead to external validity. This is mitigated to a large extent since the existing CMG tools use the fast metrics for evaluation for various reasons including the fact that even the learning based metrics have not been established to correlate significantly with human annotations better than the fast metrics. We plan to study them as future work.

%% file: 08conclusion.tex
\section{Conclusion}
\label{sec:conclusion}
%As discussed in Section~\ref{sec:experimentalsetup}, our experiments are based largely upon Tao et al.'s~\cite{sai2020survey} human evaluated dataset of size 100. However, we understand that larger the dataset, stronger the claims. Validating our results on a large scale human annotated dataset is thus an area of future work. Moreover, we believe that a better way of capturing the semantics of commit messages would result in a more accurate evaluation and increase the correlation with human evaluations. Consider a simple example pair "Update README ( )" and "Update README . md ( )". For a human annotator, these two convey the same meaning and hence receive an exact score of 1. However, metrics based on word matches would not be able to identify this pair as a perfect match.

In spite of various drawbacks, BLEU continues to be the most popular metric to be used in the evaluation of CMG models. We identify various potential factors affecting CMG evaluation. The results of our experiments show the low correlation of BLEU with human evaluated scores. With an aim of using an evaluation metric which incorporates all valid factors, a new and novel metric \textit{Log-MNEXT} is proposed. This metric has the highest correlation with human evaluations. Finally, we use this metric to compare the existing CMG models on the MCMD dataset and show that the NNGen model performs the best overall. We suggest that \textit{Log-MNEXT} should be considered as the evaluation metric for CMG evaluation in future research, instead of BLEU or its variants.

We have shared the implementation~\cite{logmnext} of \textit{Log-MNEXT} for future research.